\documentclass{article}
\usepackage{spconf,amsmath,graphicx,hyperref}
\usepackage{array}
\usepackage{amsmath,amsfonts,color}
\usepackage{algorithmic}
\usepackage[caption=false,font=normalsize,labelfont=sf,textfont=sf]{subfig}
\usepackage{textcomp}
\usepackage{stfloats}
\usepackage{url}
\usepackage{verbatim}
\usepackage{graphicx}
\usepackage{makecell}
\usepackage{bm}
\usepackage{amssymb}
 \usepackage{multirow}
\usepackage{cite}
\usepackage{booktabs}      
\usepackage{pifont}        
\usepackage{threeparttable} 
\usepackage{caption}
\usepackage{booktabs} 
\usepackage{multirow}
\usepackage{booktabs}
\usepackage{xcolor}
\usepackage{tabularx}
\usepackage{url}
\usepackage{color}
\usepackage{bm}
\usepackage{amsfonts}
\usepackage{amsmath}
\usepackage{booktabs}
\usepackage{threeparttable}
\usepackage{verbatim}

\usepackage[ruled]{algorithm2e}
\usepackage{breakurl}
\usepackage{subcaption}

\usepackage{tabularx}

\title{CaSNet: Compress-and-Send Network Based Multi-Device Speech Enhancement Model for Distributed Microphone Arrays}
\name{Chengqian Jiang, Jie Zhang, Haoyin Yan 
\thanks{Financed by National Natural Science Foundation of China (62571505), Anhui Province Major Sci. \& Tech. Research and Development Project (S2023Z20004) and Anhui Natural Science Foundation (2508085MF140).}}

\address{NERC-SLIP, University of Science and Technology of China (USTC), Hefei, China\\
  cqjiang@mail.ustc.edu.cn; jzhang6@ustc.edu.cn; hyyan@mail.ustc.edu.cn }
\begin{document}
%
\maketitle
\begin{abstract}
Distributed microphone array (DMA) is a promising next-generation platform for speech interaction, where speech enhancement (SE) is still required to improve the speech quality in noisy cases. Existing SE methods usually first gather raw waveforms at a fusion center (FC) from all devices and then design a multi-microphone model, causing high bandwidth and energy costs. In this work, we propose a \emph{Compress-and-Send Network (CaSNet)} for resource-constrained DMAs, where one microphone serves as the FC and reference. Each of other devices encodes the measured raw data into a feature matrix, which is then compressed by singular value decomposition (SVD) to produce a more compact representation. The received features at the FC are aligned via cross window query with respect to the reference, followed by neural decoding to yield spatially coherent enhanced speech. Experiments on multiple datasets show that the proposed CaSNet can save the data amount with a negligible impact on the performance compared to the uncompressed case. The reproducible code is available at https://github.com/Jokejiangv/CaSNet.
\end{abstract}
\vspace{-0.1cm}
\section{Introduction}\label{sec:intro}
\vspace{-0.1cm}

With the widespread use of sensor-equipped devices, distributed microphone array (DMA) has become a practical architecture for speech applications such as automatic speech recognition~\cite{mccowan2001robust}, teleconferencing, hearing aids~\cite{serizel2014low}, speaker localization~\cite{cobos2017survey} and speech enhancement (SE)~\cite{gannot2017consolidated}. Compared with traditional arrays in regularized shapes, the organization of DMAs brings several benefits, e.g., stronger scalability (devices are free to leave or join the array), higher signal quality, wider coverage of the acoustic scene (devices are randomly placed and some might be close to the target speaker), etc~\cite{bertrand2011applications}.

The focus of this work is on the SE task using DMAs, since SE is a necessary step to improve the signal quality in practical noisy use cases. Recent advances in deep neural networks (DNNs) have driven the design of DMA-based SE models. Many models can be applied in DMA settings, where some are DMA specified  and some are originally proposed for conventional microphone arrays. 
They are often divided into masking-based and end-to-end categories~\cite{furnon2021dnn,9054177}. The former estimates time–frequency masks (or spatial filters) in the spirit of classical beamforming and applies them to the noisy mixture to recover the target in a data-driven way~\cite{yan2025labnet,li2022embedding,kovalyov23_interspeech,furnon2021dnn}. The latter learns a direct multi-input single-output mapping, typically with convolutional neural network or recurrent neural network to model complex spatio-temporal structure~\cite{yang2023mcnet,9003849,guo2024graph,quan2024multichannel,pandey2022time}.

Most DMA-based SE models depend on aggregation of raw audio waveforms from all microphone devices to a fusion center (FC), for which high communication bandwidth and a large amount of data transmissions are required. Simple signal compression at a lower bit rate can reduce the cost, but spatiotemporal clues of the target speaker might be distorted, degrading enhancement quality and limiting downstream performance. Statistical signal processing algorithms aiming to optimize transmission bit rates or microphone subsets~\cite{zhang2019distributed,zhang2022energy} are typically restricted to stationary acoustic scenes and given second-order statistics of the entire array (this cannot be estimated in advance). This limits the applicability of elegant signal processing theories to practical DMA cases, for which the combination with popular data-driven DNN models might be a solution. More importantly, as multi-microphone signals are highly correlated and redundant~\cite{zhang2022energy}, given a prescribed SE performance it is possible to send the learned and compressed features rather than raw signals to the FC. 

In order to save the data amount, we propose a \emph{compress-and-send network (CaSNet)} for multi-device DMA-based lightweight SE. We assign one device as the FC and reference channel, and the rest as edge nodes. Each edge device converts the recorded waveform into a feature matrix, applies low-rank compression to reduce its dimensionality, and then sends the compressed feature to the FC. To capture informative characteristics of the raw signal while simultaneously reducing feature dimensionality, we employ  singular value decomposition (SVD) to generate compact representations. At the FC, the features gathered from each node are first aligned with the reference microphone using cross window query (CWQ). The aligned features are then utilized to assist the reference microphone to produce the enhanced speech signal. We evaluate the proposed CaSNet model on two public datasets and results show that with a substantial reduction in the amount of transmitted data, CaSNet maintains state-of-the-art (SOTA) SE performance.

\begin{figure*}
    \centering
    \includegraphics[width=0.95\textwidth]{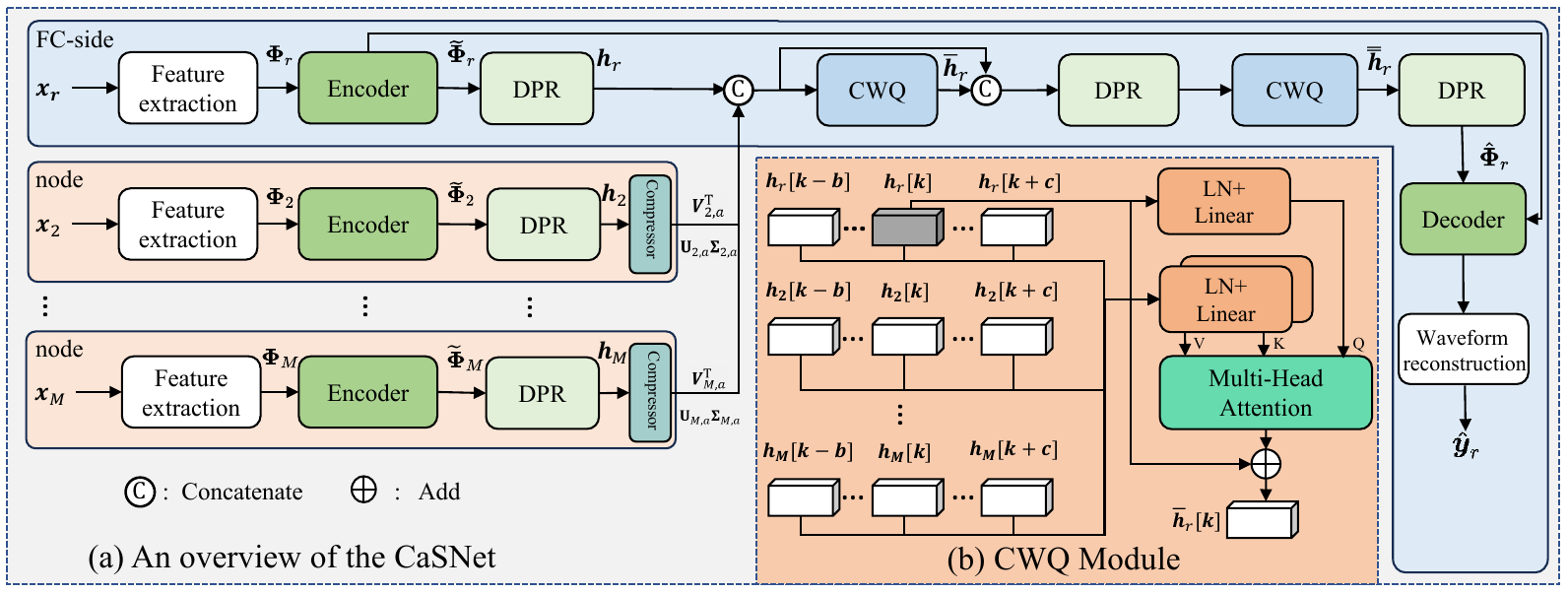}
    \caption{(a) The overall architecture of our proposed CaSNet and (b) cross-window query (CWQ) module. }
    \label{fig:architecture}
    \vspace{-1em}
\end{figure*}

\vspace{-0.15cm}
\section{CaSNet for DMA-based SE}
\vspace{-0.15cm}
In this work, we assume that each device consists of a single microphone without loss of generality. 
Given a single target speech source $\bm{s}$ and $M$ spatially separated microphones with an arbitrary geometry, the noisy observation at the $m$-th microphone is modeled as
\begin{align}
    \bm{x}_m = \bm{y}_m + \bm{n}_m = \bm{y} \circledast \bm{h}_m + \bm{n}_m, \quad m=1,2,...,M,
\end{align}
where $\circledast$ denotes convolution, $\bm{h}_m$ the room impulse response (RIR) from the source to microphone $m$, $\bm{y}_m$ the target speech component at microphone $m$, and $\bm{n}_m$ the additive noise that may include ambient noise, competing sources and late reverberation of the target.
Due to the spatial separation of microphones, the received speech at each channel undergoes different propagation delays, resulting in time misalignment. To avoid ambiguity of the target during training, we designate a reference channel $r$. Without loss of generality, we set $r=1$ in this work. Our goal is thus to estimate the target speech \( \bm{y}_r \) using noisy multichannel signals \( \{\bm{x}_m\}_{m=1}^{M} \). Some modules of CaSNet are modified from the monaural LABNet proposed by us in~\cite{yan2025labnet}. The model structure is shown in Fig.~\ref{fig:architecture}(a).

\vspace{-0.15cm}
\subsection{Microphone-side compress-and-send (CaS) design}
\vspace{-0.15cm}
For each node, we independently extract three feature maps: power-compressed magnitude spectrum, phase differences in the short-time Fourier transform (STFT) domain.
By concatenation, the feature of the $m$-th microphone is denoted as $\boldsymbol{\Phi}_m$ $\in$ $\mathbb{R}^{3 \times T \times F}$, where $T$ and $F$ are the numbers of frames and frequency bins, respectively. This is fed into an encoder that consists of several 
convolutional blocks in a U-Net structure to extract richer hierarchical representation $\tilde{\bm{\Phi}}_m$ $\in$ $\mathbb{R}^{D \times T \times F'}, \forall m$. The encoded feature is then processed by a dual-path recurrent neural network (DPR)~\cite{9054266}, which alternately models long-range dependencies along the temporal and spectral dimensions, yielding the node-level representation ${\bm{h}}_m \in \mathbb{R}^{D \times T \times F'}$. Both the encoder and DPR follow the designs in~\cite{yan2025labnet} and share parameters across all devices. 

In the compressor, we decompose $\bm{h}_m(t) $ using SVD as
\begin{align}
\bm{h}_m[t] = \mathbf{U}_m \mathbf{\Sigma}_m \mathbf{V}_m^\top \label{eq:svd_full},
\end{align}
where $(\cdot)^\top$ means vector/matrix transpose, $\mathbf{\Sigma}_m$ is a diagonal matrix containing singular values at a decreasing order as $\sigma_1\ge \sigma_2\ge\ldots\ge \sigma_D$ with $D$ being the feature dimension, and matrices $\mathbf{U}_m=[\mathbf{u}_1,\ldots,\mathbf{u}_D]$ and $\mathbf{V}_m=[\mathbf{v}_1,\ldots,\mathbf{v}_{F'}]$ contain the corresponding left and right singular vectors. Given a rank $a\leq D$, we can construct $\bm{h}_m[t] $ 
using rank-$a$ approximation as
\begin{align}
\hat{\bm{h}}_m[t] \approx \sum_{i=1}^a\sigma_i\mathbf{u}_i \mathbf{v}_i^\top, t=1,\ldots,T, \label{eq:svd_lowrank}
\end{align}
which is performed at the FC to recover feature map $\hat{\bm{h}}_m,m\in\{1,\ldots,M\} \backslash r$.
Increasing the rank can improve the approximation accuracy.
As such, for each time frame we only need to send $\mathbf{U}_{m,a}\mathbf{\Sigma}_{m,a} \in \mathbb{R}^{D \times a}$ 
and $\mathbf{V}_{m,a}^\top \in \mathbb{R}^{a \times F'}$ to the FC, instead of sending $\bm{h}_m$.   This can save the transmitted data amount compared to the uncompressed feature or raw audio data with proper choices of rank $a$.
\vspace{-0.1cm}
\subsection{FC-side cross-window query (CWQ) and decoding}
\vspace{-0.1cm}
At the FC, the reference microphone undergoes the same processing as edge nodes to produce its hidden representation $\bm{h}_r $. The CWQ module aims to alleviate temporal misalignment caused by transmission losses and asynchronous sampling rate between nodes. 
As shown in Fig.~\ref{fig:architecture}(b), we obtain each frame by short-time framing and windowing the speech signal. For the $k$-th frame of the reference channel, the hidden representation  $\bm{h}_r[k] \in \mathbb{R}^{D \times F'}$
is taken as \emph{query}, while representations 
$\{\hat{\bm{h}}_m[k-b], \ldots, \hat{\bm{h}}_m[k+c]\}$ at interval $[k-b,k+c]$ from all channels serve as \emph{keys} and \emph{values} after layer normalization and linear projection. The multi-head attention (MHA) mechanism ~\cite{NIPS2017_3f5ee243} is then applied, enabling the query frame $\bm{h}_r[k]$ to attend contextual frames rather than strict one-to-one alignment. The output of CWQ is given by
\begin{equation}
\overline{\bm{h}}_r[k] = \mathrm{MHA}\big(\bm{h}_r[k], \{\hat{\bm{h}}_m[k-b:k+c]\}\big) + \bm{h}_r[k],
\end{equation}
which can effectively compensate for delays and distortions in the transmitted features and improves the spatial coherence for subsequent decoding.

\newcolumntype{M}[1]{>{\centering\arraybackslash}m{#1}}

\begin{table*}[t]
\centering
\caption{Comparison of CaSNet with other models on both  WSJ0-WHAM! and RealMAN datasets with $M$=6 microphones.}
\small  
\begin{tabularx}{\textwidth}{@{\extracolsep{\fill}} 
  M{2.5cm} M{1.5cm} M{1.5cm} M{1.5cm} | c c c | c c c}
\toprule
\multirow{2}{*}{Method} & \multirow{2}{*}{Para.} & \multirow{2}{*}{MACs} & \multirow{2}{*}{NSA}
& \multicolumn{3}{c|}{WSJ\_WHAM!} 
& \multicolumn{3}{c}{RealMAN} \\
& & & & PESQ & STOI & COVL & PESQ & STOI & COVL \\
\midrule
Noisy & - & -  & 1 & 1.48 & 0.81 & 2.27 & 1.17 & 0.74 & 2.03 \\
Oracle MVDR & -  & - & 1 & 2.18 & 0.78 & 3.06 & 1.33 & 0.78 & 2.11 \\
FaSNet~\cite{9003849}  & 2.8M & 9.1G & 1  & 2.18 & 0.90 & 2.93 & 1.54 & 0.81 & 2.34 \\
FaSNet-TAC~\cite{9054177} & 2.3M & 11.7G  & 1 & 2.25 & 0.91 & 2.99 & 1.65 & 0.84 & 2.11 \\
DFSNet~\cite{kovalyov23_interspeech} & 549k & 1.7G & 1 & 2.01 & 0.87 & 2.79 & - & - & - \\
EaBNet~\cite{li2022embedding} & 2.8M & 7.4G & 1 & 2.76 & 0.93 & 3.64 & 2.17 & 0.91 & 3.08 \\
McNet~\cite{quan2024multichannel} & 1.9M & 30.1G & 1 & 2.82 & 0.94 & 3.67 & 2.12 & 0.90 & 3.14 \\
LABNet~\cite{yan2025labnet} & 52k & 0.316G & 1 & 2.92 & 0.93  & 3.69 & 2.34 & 0.90 & 3.14\\
\midrule
CaSNet ($a$=4) & 52k & 0.325G & 0.75 & 2.92 & 0.92  & 3.69 & 2.27 & 0.89 & 3.06\\
\bottomrule
\end{tabularx}
\label{tab:comparison3}
\end{table*}

After obtaining the enhanced reference embedding $\overline{\bm{h}}_r$, we align the features from all channels with respect to the reference. Specifically, $\overline{\bm{h}}_r$ is concatenated with node-specific feature $\bm{h}_m$, followed by a linear projection and a DPR module to produce aligned representations
The aligned features from all devices are concatenated along the channel dimension and subsequently fused by a second CWQ module, which produces a refined reference embedding $\overline{\overline{\bm{h}}}_r$.
Finally, a DPR module is applied to the fused embedding to obtain the enhanced representation $\hat{\bm{\Phi}}_r$.
To this end, the decoder can reconstruct the magnitude spectrum using $\hat{\bm{\Phi}}_r$ together with the intermediate feature maps from each layer of the reference microphone’s encoder via skip connections. Finally, the enhanced signal \( \hat{\bm{y}}_r \) is obtained by inverse STFT.

\section{Performance Evaluation}
\subsection{Experimental setup}
We evaluate our model on two datasets, where all audio is sampled or resampled to 16~kHz.

\textit{1) WSJ0-WHAM!:} A simulated dataset constructed from WSJ0 clean speech~\cite{wsj0} and WHAM! noise~\cite{wichern19_interspeech}. Each mixture contains one target speaker and 1-3 noise sources with signal-to-noise ratio (SNR) randomly sampled in $[-5, 15]$~dB. It includes 24.9 hours data for training, 2.2 hours data for validation and 1.5 hours data for testing. Training and validation involve up to 6 microphones, while testing uses up to 12 microphones with random channel selection. RIRs are generated using the image method~\cite{Allen1976ImageMF} via \texttt{gpuRIR}~\cite{Diaz_Guerra_2020}.

\textit{2) RealMAN:} A real-recorded dataset~\cite{yang2024realman} collected with a 32-channel array, containing 83.7 hours speech and 144.5 hours background noises across diverse acoustic scenes. The \texttt{ma\_speech} track is used as the target and \texttt{ma\_noise} as additive noise with SNR $\in[-5, 15]$~dB. We randomly select 6, 6 and 12 channels for training, validation and testing, respectively, 
to simulate arbitrary microphone placements.




\textbf{Implementation:} During training, we randomly select 1--6 microphones from the available channels for each mixture to simulate dynamic microphone array setup. The training utterances are cut into 4-second segments with 50\% overlap to increase sample diversity. We apply STFT with a Hann window of 512 (32 ms) samples and a hop size of 256 (16 ms)
The hyper-parameter setup of the encoder, DPR, and decoder modules in our model remains identical to those of LABNet~\cite{yan2025labnet}. The Griffin-Lim algorithm (GLA)~\cite{1164317,6701851} is applied for speech reconstruction with one iteration. For the SVD-based compressor, the rank $a$ is fixed at 4. To ensure causality, the CWQ module is configured with $b=2$ and $c=0$, while the number of attention heads is set to 4. The model is trained using the AdamW~\cite{loshchilov2019decoupled} optimizer for 100 epochs, with the gradient clipping factor set to 5.0. The learning rate starts from 5e-4 and decays at a factor of 0.98 per epoch.
\vspace{-0.2cm}
\subsection{Experimental results}
We compare the proposed CaSNet with several SOTA multichannel SE systems, including conventional beamforming baseline (Oracle MVDR), fixed-array models such as EaBNet~\cite{li2022embedding} and McNet~\cite{quan2024multichannel}, as well as adaptive array models like FaSNet~\cite{9003849}, FaSNet-TAC~\cite{9054177}, DFSNet~\cite{kovalyov23_interspeech} and LABNet. The SE performance is assessed using three widely-adopted objective metrics: perceptual evaluation of speech quality (PESQ)~\cite{rix2001pesq}, short-time objective intelligibility (STOI)~\cite{5713237}  and COVL (overall quality)~\cite{4389058}. 

Table~\ref{tab:comparison3} summarizes the overall performance of these models on the WSJ0-WHAM! and RealMAN datasets, where we also include the normalized sample amount (NSA) per microphone with respect to the case of sending the raw audio data. Note that all other models do not consider feature compression in literature, so we regard them as directly sending time samples\footnote{For compressed features or raw audio data, we use the same quantization rate per sample. NSA can thus measure the transmission cost in the case of wireless acoustic sensor network. Given $t$-seconds raw audio data and a sampling frequency of $f_s$, other models require to send $tf_s$ time-domain samples from each microphone to the FC. For CasNet, only $T(D+F')a$ samples have to be sent, which can be easily smaller than $tf_s$.}. 
Clearly, the proposed microphone-side CaS design with $a$ = 4 can reduce the data amount for collection. More importantly, CaSNet achieves competitive performance and outperforms fixed-array models (e.g., EaBNet and McNet in terms of PESQ, STOI and COVL), confirming its capacity of leveraging multichannel spatiotemporal clues. Note that the numbers of microphones used for training and testing of fixed-array models have to be the same. Meanwhile, although several modules in CaSNet originate from LABNet, the performance remains nearly unchanged compared to LABNet. This shows that CaSNet maintains a practical balance between data efficiency and SE performance. Besides, LABNet and CaSNet are much more lightweight than SOTA  models.

\newcolumntype{C}[1]{>{\centering\arraybackslash}m{#1}}

\begin{table}[t]
\centering
\caption{Evaluation of PESQ and COVL in terms of the number of microphones for LABNet and CaSNet (rank $a$=4).}
\vspace{-0.25cm}
\begin{tabularx}{\linewidth}{@{\extracolsep{\fill}} C{1.6cm} | c c | c c}
\toprule
\multirow{2}{*}{\makecell{$M$}} 
  & \multicolumn{2}{c|}{PESQ} 
  & \multicolumn{2}{c}{COVL} \\
  & LABNet & CaSNet & LABNet & CaSNet \\
\midrule
2   & 2.712 & 2.705 & 3.490 & 3.487 \\
3   & 2.790 & 2.789 & 3.569 & 3.570 \\
4   & 2.854 & 2.851 & 3.636 & 3.630 \\
5   & 2.893 & 2.891 & 3.672 & 3.669 \\
6   & 2.922 & 2.920 & 3.694 & 3.692 \\
7   & 2.945 & 2.942 & 3.722 & 3.719 \\
8   & 2.959 & 2.958 & 3.732 & 3.734 \\
9   & 2.973 & 2.969 & 3.749 & 3.745 \\
10  & 2.984 & 2.981 & 3.760 & 3.757 \\
11  & 2.989 & 2.988 & 3.766 & 3.764 \\
12  & 3.002 & 2.996 & 3.793 & 3.772 \\
\bottomrule
\end{tabularx}
\label{tab:applicability}
\vspace{-0.5em}
\vspace{-0.25cm}
\end{table}
Since CaSNet supports arbitrary numbers of input channels, we compare it with LABNet under the same configurations. Both models are trained with randomly sampled 1-6 channels and evaluated on 2-12 channels. As shown in Table~\ref{tab:applicability}, the performance of both models consistently improves as the number of input channels increases, which reflects the benefit of exploiting richer spatial cues. In all channel cases, CaSNet achieves performance comparable to LABNet, with average differences in PESQ or COVL within 0.01. This reveals that the proposed CaS design substantially reduces data overhead while preserving enhancement quality, even when the number of input microphones exceeds the training range, thereby validating the applicability of CaSNet to DMAs with arbitrary and dynamic structures.

\begin{figure}
    \centering
    \includegraphics[width=0.92\columnwidth]{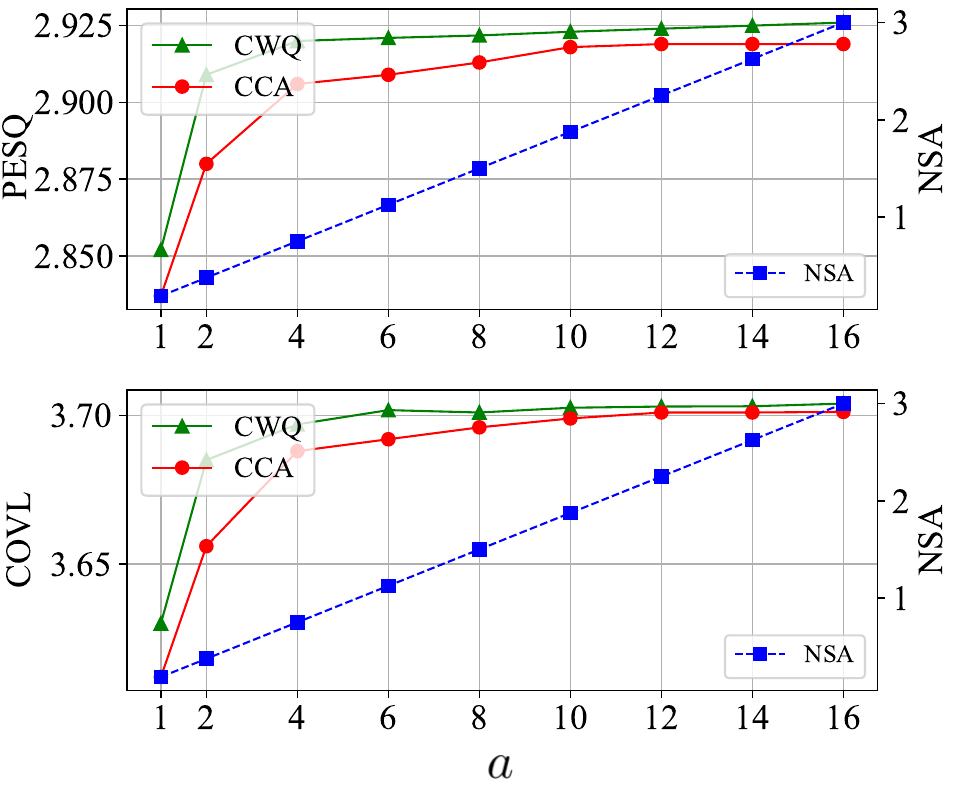}
     \vspace{-0.1cm}   
    \caption{The SE performance  and NSA in terms of rank $a\leq D=16$ with $M=6$.}
    \vspace{-0.3cm}
    \label{fig:fig1}
\end{figure}

Finally, we analyze the effects of the proposed CWQ module and the rank $a$ in Fig.~\ref{fig:fig1}. Indeed, CWQ can be replaced by cross-channel attention (CCA), similarly as LABNet. 
We see that both PESQ and COVL increase steadily with a larger $a$, since more feature information is preserved by data gathering. CWQ consistently outperforms CCA, especially in the low-rank regime ($a \leq 6$), where temporal misalignment is more serious due to over-compression. This validates the effectiveness of CWQ in the DMA-based SE framework. Note that calculations of CWQ and CCA are independent of feature compression. Increasing $a$ also causes a larger NSA and sending STFT features might cause a higher data amount than the raw case, because more eigenvectors have to be sent.
As the performance approximately saturates in case $a\ge 4$, we can simply choose $a=4$  to achieve a tradeoff between SE performance and data efficiency and incorporate the superiority of CWQ. This also reveals the data/feature redundancy in multi-microphone speech processing models~\cite{zhang2021Study}.

\vspace{-0.2cm}
\section{Conclusion}
    \vspace{-0.2cm}
In this work, we proposed the lightweight CaSNet model for DMA-based SE in an FC-node collaborative way. Edge nodes record audio signals, extract acoustic features and compress the features using SVD. The FC acts as a reference and gathers feature maps from other microphones instead of conventional raw audio data. The CWQ module is used to alleviate temporal misalignment and reconstruct spatially coherent representations. Given the same SE performance, results on public datasets clearly reveal the superiority of CaSNet in saving data amount. CaSNet also supports arbitrary numbers of microphones, i.e., scalable to practical DMA setups. Future work covers learning-based adaptive bit-rate control, sensor scheduling and joint optimization with downstream tasks.

\bibliographystyle{IEEEbib}
\bibliography{strings,refs}

\end{document}